# Impact of Oxygen on DNA Damage Distribution in 3D Genome and Its Correlation to Oxygen Enhancement Ratio after High LET Irradiation

Ankang Hu[a,b], Wanyi Zhou[a,b], Xiyu Luo[c], Rui Qiu[a,b*], Junli Li[a,b†]

[a] Department of Engineering Physics, Tsinghua University, Beijing, China
[b] Key Laboratory of Particle & Radiation Imaging, Tsinghua University, Ministry of Education, Beijing, China
[c] Institute of Fluid Physics, China Academy of Engineering Physics, Mianyang, China

## Abstract

The variation of the oxygen enhancement ratio (OER) across Linear Energy Transfer (LET) currently lacks a comprehensive mechanistic interpretation and a mechanistic model. Our earlier research revealed a significant correlation between the distribution of double-strand breaks (DSBs) within 3D genome and radiation-induced cell death, which offers valuable insights into the oxygen effect. We propose a model where the reaction of oxygen is represented as the probability of inducing DNA strand breaks. Then it is integrated into a track-structure Monte Carlo simulation to investigate the impact of oxygen on the distribution of DSBs within 3D genome. Using the parameters from our previous study, we calculate the OER values related to cell survival. Results show that the incidence ratios of clustered DSBs in a single topologically associating domain (TAD) (case 2) and DSBs in frequently-interacting TADs (case 3) under aerobic and hypoxic conditions align with the trend of the OER of cell survival across LET. Our OER curves exhibit good correspondence with experimental data. This study provides a potentially mechanistic explanation for changes in OER across LET. High-LET irradiation leads to dense ionization events, resulting in an overabundance of lesions that readily induce case 2 and case 3, which have substantially higher probabilities of cell killing than other damage patterns. This may contribute to the main mechanism governing the variation of OER for high LET. Our study further underscores the importance of the DSB distribution within 3D genome in the context of radiation-induced cell death.

## 1 Introduction

Oxygen stands as one of the most crucial physiological factors influencing the radiobiological effect. In the case of hypoxic cells, approximately three times the X-ray dose that is effective for aerobic cells is needed to achieve the same level of cell killing (*1*). The Oxygen Enhancement Ratio (OER) is defined as the ratio of the doses under hypoxic and aerobic conditions to yield the same effect. Notably, the OER diminishes as the Linear Energy Transfer (LET) increases (*2*). In the field of radiotherapy, understanding the mechanistic interpretation of the difference in OER between low and high LET irradiation is of great significance. This understanding not only sheds light on the fundamental processes underlying radiation-related cell responses but also plays a crucial role in exploring the mechanism of radiosensitizers that operate on principles similar to those of oxygen (*3*–*5*). Moreover, a mechanistic interpretation of the oxygen effect is necessary, particularly within the current framework of the FLASH effect. Here, the oxygen effect remains poorly understood, yet it may play a significant role (*6*, *7*).

[*] Corresponding author. Address: Department of Engineering Physics, Tsinghua University, Beijing, China. Email: qiurui@tsinghua.edu.cn

[†] Corresponding author. Address: Department of Engineering Physics, Tsinghua University, Beijing, China. Email: lijunli@mail.tsinghua.edu.cn

The mechanism of the radiation-oxygen effect has been investigated for many years. The oxygen-fixation hypothesis posits that oxygen reacts with DNA radicals to form peroxyl radicals (*8*, *9*). These peroxyl radicals cannot be repaired by antioxidants such as glutathione (GSH) (*10–12*). As a result, this leads to base damage or strand breaks in the DNA. Additionally, the peroxyl radicals can capture the electrons of adjacent nucleotides, thereby inducing further damage (*12–14*). The core of the hypothesis is the relationship between DSBs and the oxygen concentration. Based on these established mechanisms, researchers have developed computational models aiming to predict the OER at a specific oxygen level (*2*, *15–17*). In certain studies, the impact of oxygen is incorporated into track-structure Monte Carlo (MC) simulations. By doing so, they explore how oxygen affects DNA double-strand breaks (DSB) (*15*, *18–20*), offering a mechanistic account. These models effectively provide a mechanistic framework for OER in the context of low LET irradiation.

When it comes to the high-LET irradiation, some studies attributes the reduction of OER to the more complicated water radiolysis reactions, especially the oxygen-generation capacity, in high-LET tracks compared to low LET irradiation (*21*). The hypothesis posits that dense energy deposition by high-LET radiation triggers the multiple ionization of water molecules with the production of $\cdot HO_2$ and $H_2O_2$, which are then converted into $O_2$. Thus, even under hypoxic conditions without exogenous $O_2$, in-situ generation of $O_2$ can still occur in high-LET tracks, thereby restoring or even enhancing the efficiency of radiation-induced damage to cells (*22–24*).

According to the mechanisms from the perspective of radiochemistry, some OER models address high-LET irradiation by incorporating additional specific radiolysis processes in track-structure MC simulations(*25–30*). They introduce modifications related to the reaction between DNA radicals and oxygen or antioxidants specifically for high LET scenarios (*15*). The simulation results reveal that the OER, as defined by the number of DSBs, decreases with increasing LET (*18*). Intriguingly, this trend does not align with the OER observed in cell survival studies. In addition, some models directly utilize an empirical formula to calculate the parameters within the survival curve, thereby attempting to account for the oxygen effect during high-LET irradiation (*2*, *16*, *31*). However, these empirical methods fall short of providing a comprehensive explanation of the OER for high LET irradiation. They lack the ability to elucidate the complex underlying processes governing the OER in such high-LET situations, leaving a significant gap in our understanding of the phenomenon.

Our recent research reveals that the distribution of DSB in 3D genome exhibits a strong correlation with cell death, particularly for high LET irradiation (*17*). In this study, we incorporated oxygen into the simulation and investigated how oxygen influences the DSB distribution in the 3D genome. The results demonstrate that this oxygen-related DSB distribution in the 3D genome also has a strong correlation with the cell-survival-defined OER. Our research provides valuable perspectives for interpreting the mechanism of the OER in high LET irradiation.

# 2 Materials and Methods

## 2.1 Model for cell survival as determined by DSB distribution in the 3D genome

Our previous study found that radiation-induced cell death strongly correlates with DSB distribution in 3D genome. We proposed a hypothesis positing that when DSBs are located in DNA segments with frequent interaction, the likelihoods of incorrect repair and ultimately cell death are substantially higher than those for DSBs located within segments that interact infrequently. The proximity of two DNA segments in the genome emerges as a crucial determinant. If the segments are in close genomic proximity, there is a greater probability of small-segment structural variations occurring. On the other hand, a large genomic distance separating the segments makes the occurrence of large-segment structural variations more probable, and such large-segment variations are more prone to induce cell death.

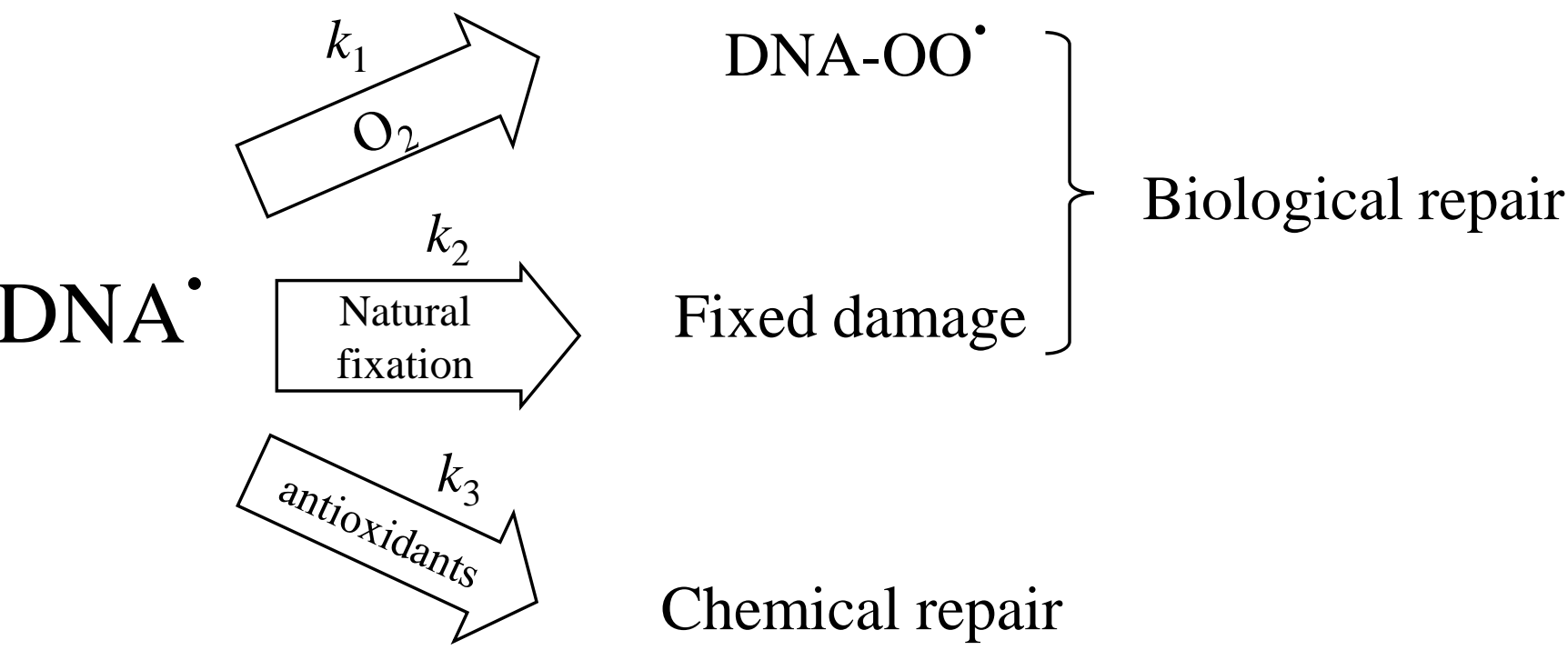

Figure 1. Diagram of DNA radical reactions

Furthermore, we established a simplified model aimed at quantitatively exploring these correlations. The distribution of DSBs within 3D genome is categorized into three distinct cases based on Topologically Associating Domains (TADs) (*32*):

- Case 1 depicts an isolated DSB scenario. In this case, a single TAD contains just one DSB, and the TADs that usually interact with this particular TAD are devoid of any DSBs.
- Case 2 is characterized by the presence of clustered DSBs within a single TAD. Here, a single TAD contains two or more DSBs.
- Case 3 pertains to DSBs in TADs that exhibit frequent interactions. Each of these TADs contains one or more DSBs, and these TADs interact frequently with one another.

For a specific dose value, the incidences of the three aforementioned cases are denoted as $n_1(D)$, $n_2(D)$ and $n_3(D)$ respectively. The probabilities of these three cases leading to cell death are $p_1$, $p_2$ and $p_3$. Consequently, the survival fraction can be computed using Equation 1.

$$-\ln S = p_1 n_1(D) + p_2 n_2(D) + p_3 n_3(D) \tag{1}$$

Our previous study indicates that p3>p2>p1. These three parameters are specific to cell types but independent of LET.

## 2.2 Mechanistic modeling of oxygen effect

Based on prior research, the impact of oxygen on radiation-induced damage primarily occurs via its reaction with DNA radicals. The DNA radical can undergo three competitive processes: (1) oxygen fixation, where it reacts with oxygen to form a peroxyl radical; (2) natural fixation, during which it transforms into damage that demands biological repair; (3) chemical repair by antioxidants such as GSH. These reactions are briefly depicted in Figure 1.

From a chemical kinetics perspective, the probability of a DNA radical transforming into a lesion can be computed using Equation 2. This formula bears a resemblance in form to the classical Alper's formula for the radiation oxygen effect (*33*).

$$p_{\text{lesion}}([\mathrm{O_2}]) = \frac{k_1 \cdot [\mathrm{O_2}] + k_2}{k_1 \cdot [\mathrm{O_2}] + k_2 + k_3 \cdot [\text{antioxidants}]} \tag{2}$$

Where $k_1$, $k_2$ and $k_3$ represent reaction rate constants, while $[O_2]$ and [antioxidants] denote the concentrations of oxygen and antioxidants. The oxygen effect can largely be ascribed to its influence on the probability of DNA radicals converting into lesions. For conventional dose-rate irradiations, the reaction rate constants and the concentrations of reactants remain unaffected by the LET. As a result, Equation 2 also holds for high-LET irradiations. Consequently, in our study, the oxygen-related probability is set to be identical at the same oxygen level across all LET values.

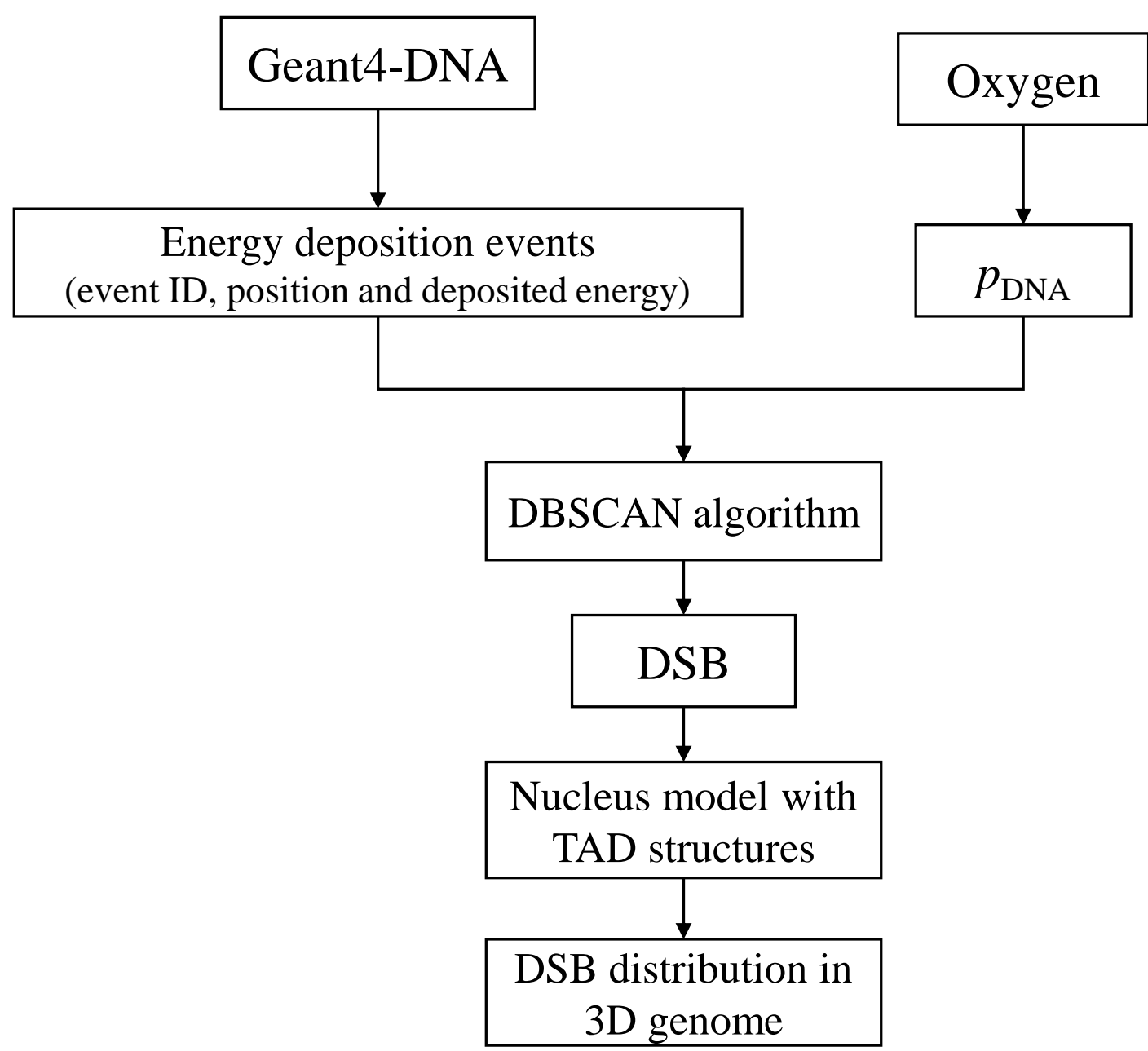

Figure 2. Diagram of the simulation process

## 2.3 Track-structure Monte Carlo simulation

In our study, we utilize Geant4-DNA to simulate energy deposition events. Subsequently, these energy deposition events are analyzed to determine the distribution of DSBs within the 3D genome of the nucleus model incorporating TAD structures. When analyzing DNA damage, the oxygen effect is factored in by considering the probability of an energy deposition event being converted into a strand break ($p_{DNA}$). The simulation processes are depicted in Figure 2, and detailed accounts are presented in the sub-sections that follow.

### 2.3.1 Model of nucleus

For our study, we employ a nuclear model of the IMR90 cell. This model, developed by Ingram et al. (*33*). using the code G-NOME based on Hi-C data, defines TAD structures. The nucleus in this model has a radius of 5.0 µm and an occupancy volume of 0.15. TADs are represented as spherical units, with their positions and radii determined by G-NOME. In total, there are 15,282 TADs, encapsulating 6.07 Gbp of DNA. The mean radius of the TAD spheres is 89.12 nm, with a standard deviation of 41.29 nm, and the mean DNA content per TAD is 397 kbp, having a standard deviation of 840 kbp.

Moreover, the model furnishes details about TADs that frequently interact with a particular TAD. The criteria for identifying such frequently-interacting TADs were developed by Chrom3D through a statistical testing method. Significant interactions are pinpointed using the NCHG module within Chrom3D (*34*, *35*). This module calculates a $P$ value based on the probability of observing a specific number of contacts, considering the total number of contacts for both involved regions and the overall number of contacts, by applying a non-central hypergeometric distribution. Interactions are then selected with a false-discovery rate set at 1%.

We utilized a single nucleus in the simulation because, for all G-NOME-generated nucleus models with the same Hi-C data, the sizes of TADs and the relative contacts among them remain consistent. In each simulation, we randomly rotated the coordinates of energy deposition events by a certain angle. This was done to simulate the diverse angles at which the cell might be exposed to radiation.

### 2.3.2 Physics simulation

We conduct the physics simulations using Geant4 version 11.2, with the physics list set as G4EmDNAPhysics_option2 (*36*). The simulation world is defined as a water cube having an edge length of 120 μm, and the nuclear model is placed precisely at the center of this cube. The ion source is modeled as a plane located 5.5 μm away from the cube's center, ensuring that all the TAD spheres are positioned in front of the source. The starting points of the particles are uniformly and randomly sampled from this plane, and their directions are configured to be parallel to the normal vector of the plane.

For the simulation of each particle energy, we establish distinct physics simulation groups to approximate a dose range spanning from 1.0 to 10.0 Gy. For every one of these physics simulation groups, the physics simulations are executed 10 times, with a fixed number of simulated particles in each execution. The number of particles is determined based on the dose delivered to a sphere with a radius of 5 μm, which is positioned at the center of the simulation space. Specifically, we select the particle number corresponding to a dose close to 1.0 Gy.

### 2.3.3 Simulation of oxygen effect and analysis of DNA damage

After the physics simulation, the energy deposition events are initially filtered based on energy. The probability of an event leading to potential lesions increases linearly, starting from 0 at 5 eV and reaching 1 at 37.5 eV. Subsequently, all potential lesions within the TAD spheres are randomly retained with a probability ($p_{\mathrm{DNA}}$). These retained lesions are then randomly assigned to either strand 1 or strand 2 of the DNA double helix, each with an equal probability. DSBs are defined as two or more strand breaks that occur on opposite strands and are separated by a distance of 3.2 nm or less, which is roughly equivalent to 10 base pairs. All parameters in this setup, except for $p_{\mathrm{DNA}}$, are consistent with those used in our previous study. The Density-Based Spatial Clustering of Applications with Noise (DBSCAN) algorithm is utilized to analyze the distribution and clustering of DSBs. Following this, the incidences of three cases are counted for subsequent analysis (*32*, *34*).

The value of $p_{\mathrm{DNA}}$ is governed by two key factors: the probability of an energy deposition event striking the DNA strand ($p_{\mathrm{hit}}$) and the probability of that event transforming into a strand break ($p_{\mathrm{lesion}}$), which is dictated by the oxygen effect. This value can be computed using Equation 3.

$$p_{\mathrm{DNA}} = p_{\mathrm{lesion}}([\mathrm{O_2}]) \cdot p_{\mathrm{hit}} \tag{3}$$

For each physics simulation run, the analysis of DSB is performed 100 times. In each instance of the DSB analysis, the coordinates of the energy deposition events are rotated by a random angle. The results of the DSB analysis provided estimates of the average incidences of the cases induced by the specific dose (as defined by the centered sphere). We assume that these incidences followed a Poisson distribution, in which the variance is equal to the mathematical expectation. As a result, our results inherently encapsulated information regarding the associated variability.

### 2.3.4 Calculating OER for irradiation with different LETs

We perform simulations using 0.9 MeV electrons to represent low LET irradiation and employ protons with energies ranging from 0.4 to 66.46 MeV, He-3 ions with energies from 0.8 to 23.0 MeV/u, and C-12 ions with energies from 2.98 to 209.46 MeV/u to study high LET irradiation scenarios.

For each specific type of particle, we systematically assigned values to $p_{\mathrm{DNA}}$ within the range of 0.04 to 0.22. This is done to investigate the impact of oxygen on the relevant processes. Additionally, we utilize the values of p1, p2, and p3 that were previously obtained from our research on cell lines such as V79, HSG, T1 and H460 (*32*, *37*, *38*). These values are then employed to calculate the OER.

# 3 Results and Discussions

## 3.1 Influence of $p_{\mathrm{DNA}}$ on damage distribution in 3D genome

The yields of single-strand breaks (SSBs), DSBs, and the incidences of the three cases are presented in Figure 3. Specifically, the results for electrons and C-12 ions with LET values of 20 and 75 keV/μm are shown to exemplify the outcomes across different LET levels. The incidences of case 1 to case 3 are calculated at the dose of 4 Gy.

For low LET irradiation, the SSB yield has a linear relationship with $p_{\mathrm{DNA}}$. This is because the probability of a strand break occurring is determined by $p_{\mathrm{DNA}}$, and the DSB yield is relatively low under such conditions. In contrast, the curves for DSBs and the three cases (case 1, case 2, and case 3) are non-linear. These cases represent complex damage scenarios involving multiple strand breaks. When it comes to high LET irradiation, the curves are even more intricate compared to low LET irradiation. DSBs induced by high LET irradiation typically consist of multiple strand breaks, and both case 2 and case 3 involve multiple DSBs. These factors contribute to the formation of the complex non-linear curves observed for high-LET irradiation.

The classical Alper's formula for the radiation oxygen effect is effective in describing the incidences of strand breaks as it shows a linear relationship with $p_{\mathrm{DNA}}$. However, for complex damage scenarios such as DSBs and more complex cases, the non-linear curves deviate from what Alper's formula would predict. Nevertheless, if the change of $p_{\mathrm{DNA}}$ caused by oxygen is not substantial, Alper's formula can still serve as a reasonably good linear approximation.

## 3.2 Impact of oxygen on damage distribution for different LET

In our prior research, $p_{\mathrm{DNA}} = 0.14$ was used to represent the aerobic condition. According to experimental results of OER, in the present study, we set $p_{\mathrm{DNA}} = 0.09$ as the parameter representing the hypoxic condition.

Figure 4 showcases the ratios of incidences between $p_{\mathrm{DNA}} = 0.14$ and $p_{\mathrm{DNA}} = 0.09$, thereby demonstrating the influence of oxygen on SSB, DSB, case 1, case 2, and case 3 for different LET. The incidences of case 1, case 2, and case 3 are calculated at a dose of 4 Gy. Specifically, the incidences of Case 2 and Case 3 are differentiated based on whether they are induced by single particles or multiple particles. This figure enables a clear visualization of how the LET impacts oxygen-related damage.

Our results demonstrate that as the LET increases, the yield ratios of SSBs and DSBs, along with the incidence ratios of case 2 and case 3, decline. When contrasted with low-LET irradiation, high-LET irradiation generates dense ionization. In high-LET scenarios, a single DSB encompasses more extensive DNA damage. Additionally, both case 2 and case 3 entail a greater number of DSBs, resulting in an overabundance of lesions. Even if some of these lesions undergo chemical repair, the remaining ones are sufficient to result in DSBs, as well as the occurrence of case 2 and case 3. However, the limited impact changes of SSB and DSB are insufficient to account for the trend observed in the OER of cell survival. As LET approaches 100 keV/μm, the OER of cell survival drops to approximately 1.0. It indicates that other factors may play a crucial role in modulating the oxygen effect at high LET values.

The incidence ratio curves of case 2 and case 3 across different LET values exhibit a trend remarkably similar to that of the OER of the survival fraction. This similarity strongly indicates a significant correlation between the incidences of case 2 and case 3 and the OER. Findings from our previous study revealed that the probabilities of cell death induced by case 2 and case 3 are substantially higher than those induced by isolated DSBs. This evidence further supports our explanation that the decline in the incidences of case 2 and case 3 with increasing LET represents a main factor contributing to the observed decrease in the OER.

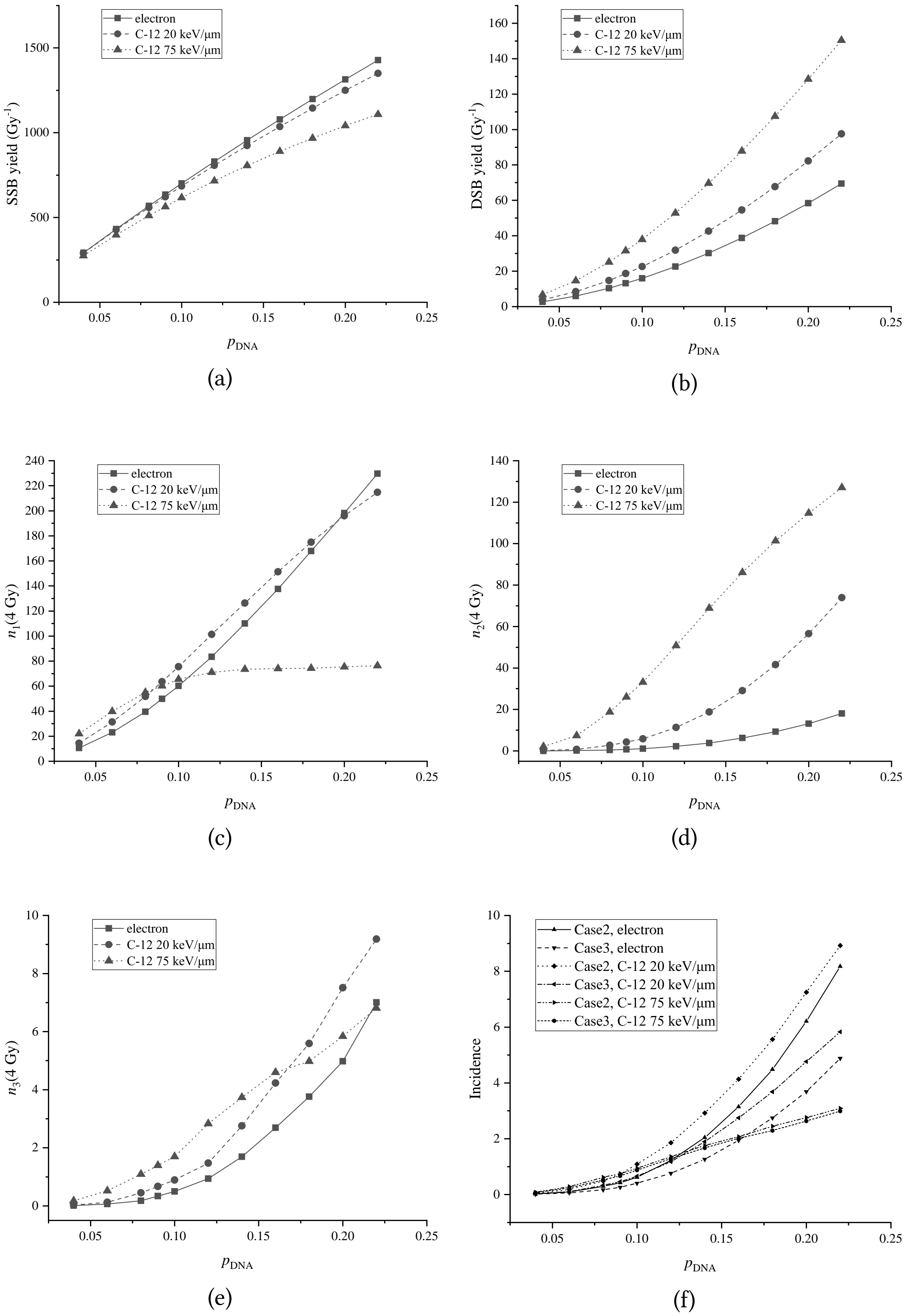

Figure 3. Yields of (a)SSB, (b)DSB and incidences of (c)case 1, (d)case2, (e)case 3 and (f)case 2 and case 3 induced by multiple particles at 4 Gy for different $p_{\mathrm{DNA}}$ and LET

## 3.3 OER for different LET

Leveraging the probabilities of cell death induced by the three cases—findings from our previous study, we computed the OER associated with the $\alpha$ parameter ($\mathrm{OER}_{\alpha}$) and the OER at the 10% survival

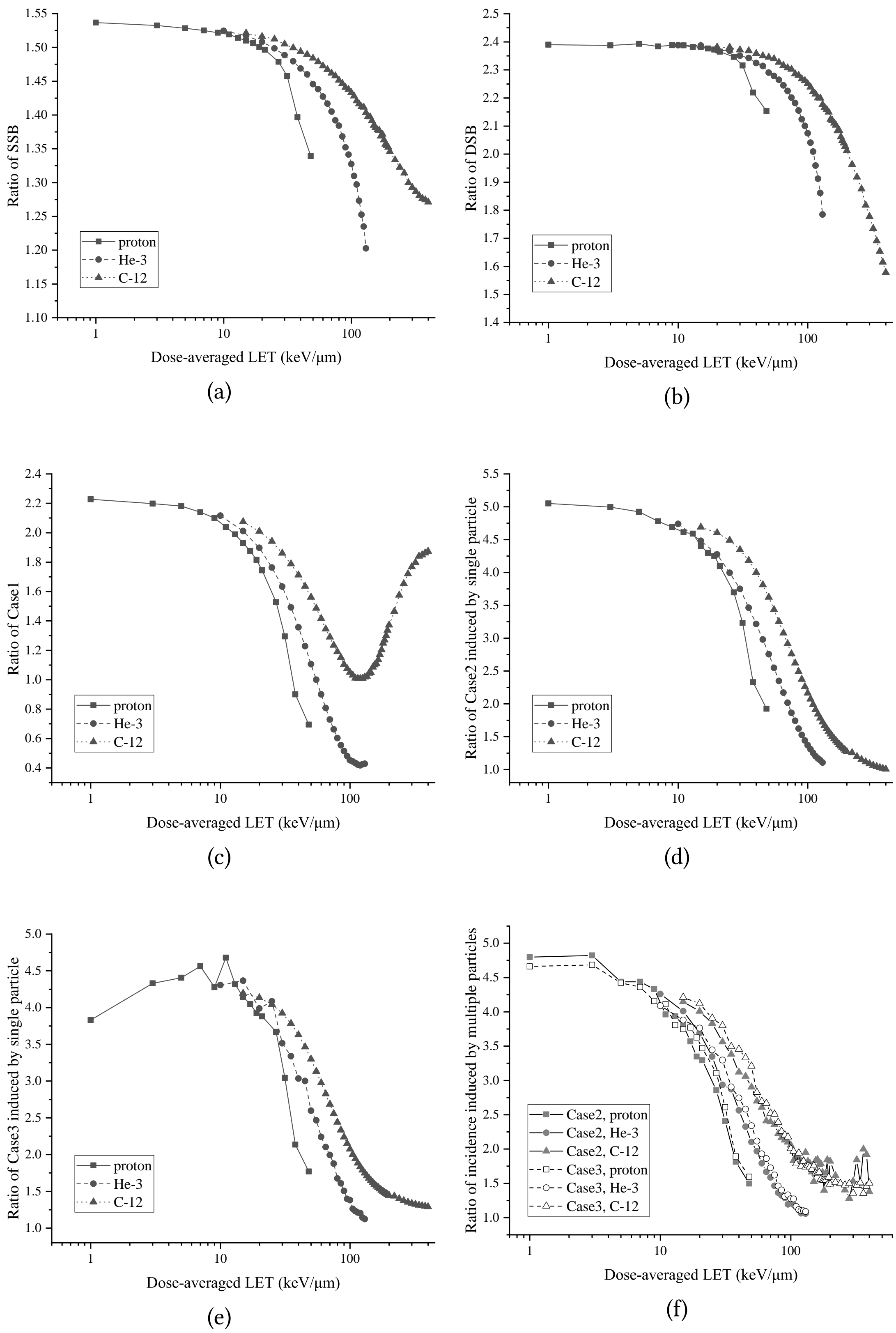

Figure 4. Ratios of incidences between $p_{\text{DNA}} = 0.14$ and $p_{\text{DNA}} = 0.09$ for different LET: (a)SSB, (b)DSB, (c)case1, (d)case2 and (e)case3 induced by single particle and (f)case2 and case3 induced by multiple particles

fraction ($\text{OER}_{10}$). The results of $\text{OER}_{\alpha}$ are presented in Figure 5. In Figure 6, the results of the $\text{OER}_{10}$

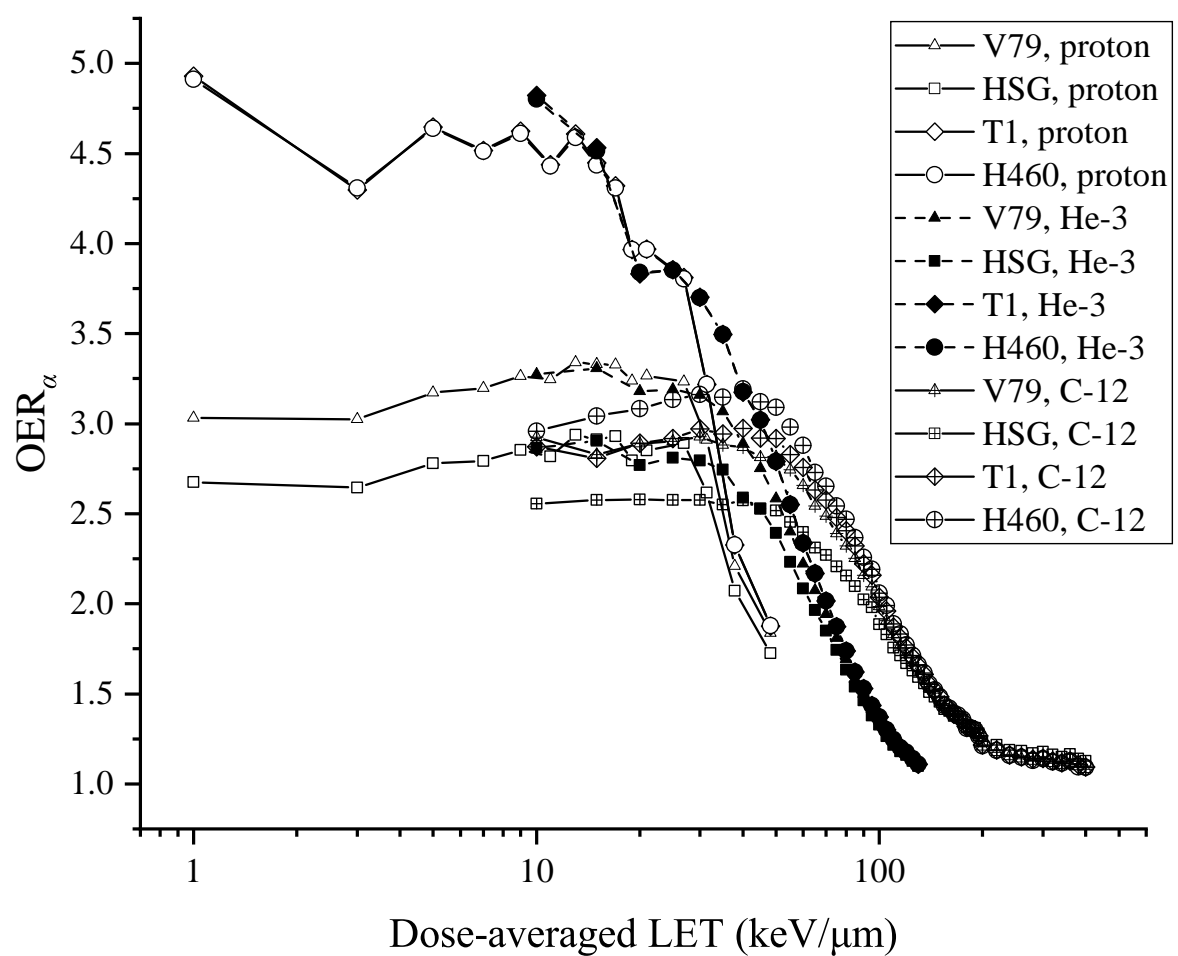

Figure 5. OER$_\alpha$ across different LET predicted by our model

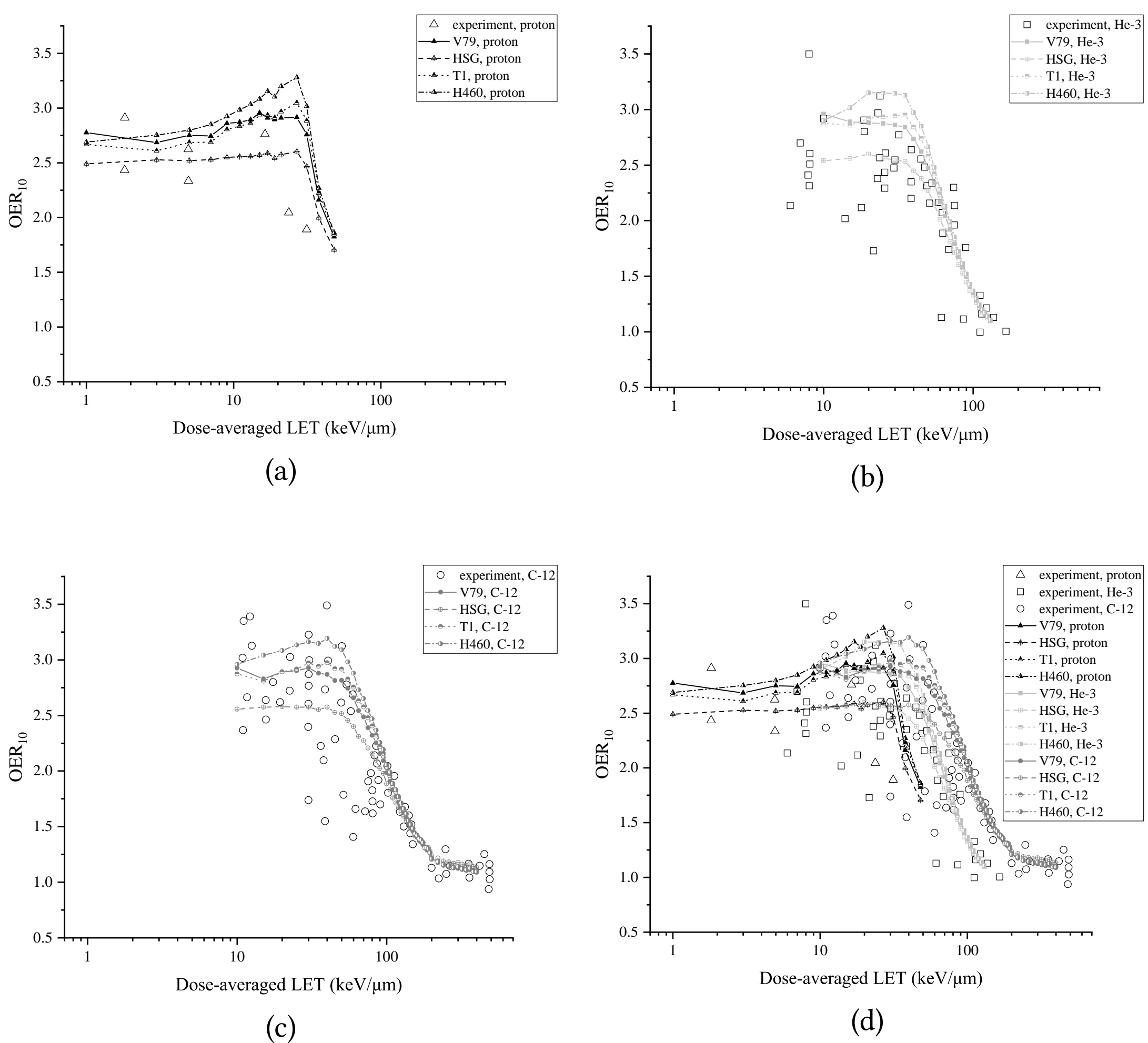

Figure 6. Comparison of OER at 10% survival fraction across different LET between experimental data and predictions from our model: (a) proton, (b)He-3 ion, (c) C-12 ion and (d)all data

are presented and juxtaposed against the experimental data points that were collated and summarized by Wenzl and Wilkens (*2*).

Our curves exhibit a trend closely resembling that of the experimental data. However, conducting a more in-depth quantitative comparison is unfeasible due to the significant scatter in the experimental data points. Our results suggest that the $OER_{10}$ demonstrates a marginal increase at LET values in the range of tens of keV/μm. In this same LET range, the $OER_{\alpha}$ remains relatively constant. Nevertheless, the $\alpha/\beta$ ratio increases, causing $OER_{10}$ to gradually converge towards $OER_{\alpha}$. Given that $OER_{\alpha}$ is consistently higher than $OER_{10}$, this difference might account for the observed slight increase in $OER_{10}$.

## 3.4 Discussion on the mechanism of OER for high LET irradiation

Currently, the mechanistic understanding underlying models of the OER across different LET values remains severely inadequate. The majority of existing models rely on empirical parameters derived from direct fitting of cell-survival data. Although some models attempt to incorporate modifications related to the reactions between DNA radicals and oxygen or antioxidants for high-LET conditions, these adjustments are still fundamentally empirical. From a chemical perspective, LET should influence the reactions of both oxygen and GSH with DNA radicals to an equal extent. The modifications introduced in these models lack support from established chemical principles.

In general, the decline in the OER as LET increases is attributed to dense ionization, which generates an overabundance of lesions. Even though some of these lesions undergo chemical repair, the remaining ones are still sufficient to cause lethal damage. However, there are unresolved questions regarding the physical nature of lethal lesions. The scale of such lesions should align with the trend of damage distribution across different LET values. Understanding the mechanistic basis of OER variation with LET is part of explaining the mechanism of cell death as a function of LET. To comprehensively address both, it is essential to identify the physical nature of lethal lesions.

The ratio curve of DSB yields between hypoxic and aerobic conditions across different LET values does not align with the OER curve. Similarly, the DSB yield curve fails to match the relative biological effectiveness (RBE) curve. This discrepancy suggests that a larger-scale genomic structure might play a more crucial role in radiation-induced cellular responses. Findings from our previous research indicate that the distribution of DSBs within the 3D genome, particularly the presence of multiple DSBs within a TAD or between frequently interacting TADs, exhibits a strong correlation with cell death. The current study further reveals a significant correlation between this DSB distribution pattern and the OER across various LET values. Collectively, these results underscore the importance of DSB distribution as a key determinant of radiation-induced cell death. Building upon the hypothesis that cell death is intricately linked to the distribution of DSB within the 3D genome, our study presents a potentially novel mechanistic interpretation of the radiation oxygen effect across different LET values. By highlighting this connection, we offer new insights into how the presence of oxygen modulates the impact of radiation on cells at the genomic scale, thereby contributing to a more comprehensive understanding of radiation-induced cellular responses. The FLASH effect is currently a frontier and hotspot in the field of radiotherapy. The oxygen effect may play a significant role therein, and a mechanistic understanding of the oxygen effect will also facilitate the comprehension of the mechanism behind the FLASH effect. It should be noted that while we consider the explanation proposed in this study to be the primary focus, other factors also contribute to the reduced OER observed in high LET irradiation. These include the increased oxygen yield generated by multiple ionization processes induced by high LET irradiation, although this yield exerts only a minor impact on the oxygen effect (*22*). Further detailed investigations into these aspects will be necessary in future studies.

## 3.5 Limitations

This study employs the same simulation approach as our prior research. The track-structure Monte MC simulation, the nuclear model, and the DBSCAN algorithm all introduce uncertainties. These factors remain the primary limitations of this study. Furthermore, we simplified the reactions of oxygen and antioxidants. We modeled them as the probability of an energy deposition event leading to a strand break. The detailed impacts of chemical reactions may have been overlooked. This simplification might overlook details. In particular, oxygen generated through the multiple ionization of water molecules increases with rising LET. Nevertheless, according to the findings of Gervais et al. (*22*), the oxygen yield is less than $2 \times 10^{-8}$ M/Gy—a level that exerts only a minor influence on the core mechanistic explanation of the oxygen effect proposed in our study.

The experimental data points of the OER at high LET are highly dispersed, rendering a quantitative comparison with the experimental data unfeasible. The main point of our study is to elucidate the influence of oxygen on the distribution of DNA damage within the 3D genome and to establish its correlation with the OER curve across varying LET values. Presently, due to the scatter in the data, a detailed and quantitative model of the OER as a function of LET remains elusive. It will be feasible to formulate a comprehensive and numerically precise model of the OER in relation to LET once more experimental findings validate the hypothesis from our earlier study.

# 4 Conclusion

In this study, we model the reaction of oxygen as the probability of inducing DNA strand breaks. This model is integrated into a track-structure MC simulation, enabling us to explore how oxygen impacts the distribution of DSBs within the 3D genome. Our results demonstrate that the incidence ratios of clustered DSBs within a single TAD (case 2) and DSBs in frequently-interacting TADs (case 3) under aerobic and hypoxic conditions closely mirror the trend of the OER of cell survival across different LET values. Leveraging the hypothesis and parameters established in our previous research, we calculate OER values of cell survival. The resulting OER curves exhibit good correspondence with experimental data.

Our work offers a potentially novel mechanistic explanation for the variation of OER across different LET levels. High-LET irradiation generates dense ionization, resulting in an overabundance of lesions that readily induce case 2 and case 3. The probabilities of cell death induced by case 2 and case 3 are significantly higher than those associated with other damage patterns. This finding potentially represents a main mechanism underlying the variation of OER across LET.

Moreover, our study further underscores the importance of DSB distribution within the 3D genome in radiation-induced cell death. Our research not only contributes to the fundamental understanding of radiation biology but also holds potential implications for optimizing radiotherapy and related applications.

# Acknowledgments

This work was supported by the National Key Research and Development Program of China (Grant No. 2024YFA1014103), the National Natural Science Foundation of China (Grant No. 12405359 and U2167209) and the Postdoctoral Fellowship Program of Chinese Postdoctoral Science Foundation under Grant Number GZB20240342.

# Disclosure of interest

The authors report no conflict of interest.